\documentclass[twocolumn,showpacs,preprintnumbers,amsmath,amssymb]{revtex4}
\usepackage{graphicx} % Include figure files
\usepackage{dcolumn} % Align table columns on decimal point
\usepackage{bm} % bold math
\usepackage{subfigure}

\begin{document}

%\preprint{APS/xxx-xxx}

\title{Thermoelectric phenomena in a quantum dot asymmetrically coupled to 
       external leads}

\author{M. Krawiec}
 \email{krawiec@kft.umcs.lublin.pl}
\author{K. I. Wysoki\'{n}ski}
 \email{karol@tytan.umcs.lublin.pl}
\affiliation{Institute of Physics and Nanotechnology Center, 
             M. Curie-Sk\l odowska University, Pl. M. Curie-Sk\l odowskiej 1,
	     20-031 Lublin, Poland}

\date{\today}

\begin{abstract}
We study thermoelectric phenomena in a system consisting of strongly correlated 
quantum dot coupled to external leads in the Kondo regime. We calculate linear 
and nonlinear electrical and thermal conductance and thermopower of the 
quantum dot and discuss the role of asymmetry in the couplings to external 
electrodes. In the linear regime electrical and thermal conductances are 
modified, while thermopower remains unchanged. In the nonlinear regime the 
Kondo resonance in differential conductance develops at non-zero source-drain 
voltage, which has important consequences on thermoelectric properties of the 
system and the thermopower starts to depend on the asymmetry. We also discuss 
Wiedemann-Franz relation, thermoelectric figure of merit and validity of the 
Mott formula for thermopower.
\end{abstract}

\maketitle

%%%%%%%%%%%%%%%%%%%%%%%%%%%%%%%%%%%%%%%%%%%%%%%%%%%%%%%%%%%%%%%%%%%%%%%%%%%%%%

\section{\label{introduction} Introduction}

The Kondo effect, discovered more than seventy years ago in metals with small 
amounts of magnetic impurities and explained more than forty years ago 
\cite{Kondo}, is now a prime example of many body phenomena in a wide class of 
correlated electron systems \cite{Hewson}. Due to recent advances in 
nanotechnology it became possible to fabricate artificial structures, where the 
Kondo effect can be studied systematically. Probably the best control over the 
isolated spin can be achieved in quantum dot (QD) systems. Originally, the 
Kondo effect in quantum dots was predicted theoretically in late eighties 
\cite{Glazman,Ng,Kawabata} and later confirmed in series of beautiful 
experiments \cite{Goldhaber,Cronenwett,Schmid,Simmel,Sasaki,Schmid_1,Buitelaar,
Pasupathy,Nygard}.

The origin of the Kondo effect, both in alloys and in quantum dots comes from 
the formation of a singlet state between a localized spin and free electrons. 
This takes place at temperatures lower than the Kondo temperature $T_K$ and 
manifests itself in increasing conductance ($G$) of the system. The increase of 
$G$ is a result of the formation of a many body Kondo or Abrikosov-Suhl 
resonance at the Fermi energy.

The experiments have confirmed the validity of the theoretical picture but they
also discovered phenomena which require new theoretical ideas. These include
observation of the Kondo resonance at non-zero source-drain voltage
\cite{Schmid,Simmel}, absence of even-odd parity effects expected for these
systems \cite{Schmid_1}, observation of the singlet-triplet transition in a
magnetic field \cite{Sasaki}, splitting of the Kondo resonance due to the
ferromagnetism in the leads \cite{Pasupathy} or the interplay between the Kondo
effect and superconductivity in carbon nanotube quantum dots coupled to 
superconducting leads \cite{Buitelaar}.

In the present paper we shall focus our attention on the nonequilibrium Kondo 
effect (i.e. the Kondo effect at non-zero source drain voltage) in 
asymmetrically coupled quantum dots. In our previous work \cite{MK_1} we have 
found that the asymmetry in the couplings to the leads is the main reason 
responsible for the nonequilibrium Kondo effect. The effects of non-symmetric 
coupling have also been discussed in Refs. 
\cite{Babic,Rosch,Kehrein,Sanchez,Swirkowicz}. Here we present a more 
systematic study on the role of the asymmetry in the couplings to the leads in 
thermoelectric and transport properties of quantum dot system. Thermoelectric 
properties (thermopower and thermal conductance) in the Kondo regime have 
already been investigated in the quantum dots coupled to the normal 
\cite{Boese,Dong,Kim} and the ferromagnetic \cite{MK_2,MK_3} leads, however the 
role of the asymmetry in the couplings has not been discussed so far. Thus we 
shall concentrate on the conductance, thermal conductance, thermopower and 
related quantities such as thermoelectric figure of merit which directly 
provides the information on the usefulness of the system for applications and 
Wiedemann-Franz ratio which signals breakdown of the Fermi liquid state when 
its normalized value differs from 1. 

In order to calculate those quantities we use non-crossing approximation (NCA)
\cite{Keiter}, which is a widely accepted and reliable technique to study the 
equilibrium and nonequilibrium Kondo effect in quantum dots
\cite{Wingreen,Costi,Hettler,MK_1}. Although this method is known to give
non-Fermi liquid ground state, it remains valid down to temperatures below
$T_K$ \cite{Bickers,Cox}. 

The paper is organized as follows. In Sec. \ref{theory} we introduce the model
and discuss some aspects of our procedure. The results of our calculations
regarding the electrical and thermal transport in linear and nonlinear regime 
are presented and discussed in Secs. \ref{linear} and \ref{nonlinear}, 
respectively. Finally, conclusions are given in Sec. \ref{conclusions}.

%%%%%%%%%%%%%%%%%%%%%%%%%%%%%%%%%%%%%%%%%%%%%%%%%%%%%%%%%%%%%%%%%%%%%%%%%%%%%%

\section{\label{theory} The model and approach}

The system we are studying consists of a quantum dot with a single energy level 
coupled to external electrodes. We describe it by the single impurity Anderson 
model \cite{Anderson} with  very  strong on-dot Coulomb repulsion 
($U \rightarrow \infty$) and use slave boson representation  \cite{Coleman}, in 
which the real on-dot electron operator $d_{\sigma}$ is replaced by the product 
of boson $b$ and fermion operators $f_{\sigma}$ ($d_{\sigma} = b^+ f_{\sigma}$) 
subject to the constraint 
$b^+b+\sum_{\sigma}f^+_{\sigma}f_{\sigma}=1$. The resulting Hamiltonian reads
\begin{eqnarray}
H = \sum_{\lambda {\bf k} \sigma} \epsilon_{\lambda {\bf k}} 
    c^+_{\lambda {\bf k} \sigma} c_{\lambda {\bf k} \sigma} +
    \varepsilon_d \sum_{\sigma} f^+_{\sigma} f_{\sigma} +
\nonumber \\
    \sum_{\lambda {\bf k}} \left(V_{\lambda {\bf k}} 
    c^+_{\lambda {\bf k} \sigma} b^+ f_{\sigma} + H. c. \right),
\label{Hamilt}
\end{eqnarray}
 where $\lambda = L$ ($R$) denotes left (right) lead, 
$c^+_{\lambda {\bf k} \sigma}$ ($c_{\lambda {\bf k} \sigma}$) is the creation
(annihilation) operator for a conduction electron with the wave vector 
${\bf k}$, spin $\sigma$ in the lead $\lambda$ and $V_{\lambda {\bf k}}$ is the 
hybridization parameter between localized electron on the dot with energy 
$\varepsilon_d$ and conduction electron of energy $\epsilon_{\lambda {\bf k}}$ 
in the lead $\lambda$. 

In order to calculate electrical current $J_{e\lambda}$ and energy flux
$J_{E\lambda}$ flowing from the lead $\lambda$ to the central region we follow
standard derivation \cite{Haug} and express all the currents in terms of 
Keldysh Green functions \cite{Keldysh}. Moreover, we use relation 
$J_{Q\lambda} = J_{E\lambda} - \mu_{\lambda} J_{e \lambda}$ for thermal flux 
$J_{Q\lambda}$, so the resulting expressions are
\begin{eqnarray}
J_{e \lambda} = \frac{i e}{\hbar} \sum_{\sigma} \int^{\infty}_{-\infty}
\frac{d\omega}{2 \pi} \Gamma_{\lambda}(\omega) 
[ G^<_{\sigma}(\omega) + 
2 i f_{\lambda}(\omega) {\rm Im} G^r_{\sigma}(\omega) ]
\label{p_currGF}
\end{eqnarray}
\begin{eqnarray}
J_{Q \lambda} = \frac{i}{\hbar} \sum_{\sigma} \int^{\infty}_{-\infty}
\frac{d\omega}{2 \pi} \Gamma_{\lambda}(\omega) (\omega - 
\mu_{\lambda}) 
[ G^<_{\sigma}(\omega) \nonumber \\
+ 2 i f_{\lambda}(\omega) {\rm Im} G^r_{\sigma}(\omega) ] ,
\label{h_currGF}
\end{eqnarray}
where  $\Gamma_{\lambda}(\omega) = 2 \pi \sum_{\bf k} 
|V_{\lambda {\bf k}}|^2 \delta(\omega - \epsilon_{\lambda {\bf k}})$ is the 
coupling of the dot to the lead $\lambda$, and $G^r_{\sigma}(\omega)$ is the 
time Fourier transform of retarded Green function (GF) $G^r_{\sigma}(t,t') = 
i \theta(t-t') 
\langle [b^+(t) f_{\sigma}(t), f^+_{\sigma}(t') b(t')]_+ \rangle$ and
$G^<_{\sigma}(\omega) = 
i \langle f^+_{\sigma}(t') b(t') b^+(t) f_{\sigma}(t) \rangle$ is the Fourier
transform of lesser Keldysh GF \cite{Keldysh}. 
$f_{\lambda}(\omega) = 
(\exp{(\frac{\omega - \mu_{\lambda}}{k_B T_{\lambda}})}+1)^{-1}$ is the
Fermi distribution function in the lead $\lambda$ with chemical potential
$\mu_{\lambda}$ and temperature $T_{\lambda}$.

In order to calculate lesser GF $G^<_{\sigma}(\omega)$ we use widely accepted 
Ng ansatz \cite{Ng_1}, as in the presence of both the on-dot Coulomb 
interaction and tunneling between the QD and the leads, it is not possible to 
calculate $G^<_{\sigma}(\omega)$ exactly. In this approach one assumes that 
full interacting lesser self-energy is proportional to the noninteracting one, 
and the resulting lesser self-energy is expressed in terms of retarded 
interacting and noninteracting self-energies. This approach has three 
advantages, (i) it is exact in nonequilibrium for $U = 0$, (ii) it is exact in 
equilibrium for any $U$, and (iii) it satisfies the continuity equation 
$J_L = - J_R$ in the steady state \cite{Ng_1}. In the present case it yields
\begin{eqnarray}
J_e = -\frac{2 e}{\hbar} \sum_{\sigma} \int^{\infty}_{-\infty}
\frac{d\omega}{2\pi} \tilde \Gamma(\omega) 
[f_L(\omega) - f_R(\omega)] {\rm Im} G^r_{\sigma}(\omega)
\label{p_curr}
\end{eqnarray}
\begin{eqnarray}
J_Q = -\frac{2}{\hbar} \sum_{\sigma} \int^{\infty}_{-\infty}
\frac{d\omega}{2\pi} \tilde \Gamma(\omega) (\omega - eV) \nonumber \\
\times [f_L(\omega) - f_R(\omega)] {\rm Im} G^r_{\sigma}(\omega) ,
\label{h_curr}
\end{eqnarray}
where $\tilde \Gamma = \Gamma_L \Gamma_R /(\Gamma_L + \Gamma_R)$, and 
$eV = \mu_L - \mu_R$. The on-dot retarded GF $G^r_{\sigma}(\omega)$ is 
calculated within NCA in terms of boson and fermion propagators, which are 
expressed by the coupled integral equations \cite{Wingreen,Costi,Hettler,MK_1}. 
In the numerical calculations we have used Lorentzian bands in the electrodes 
of width $D = 100 \Gamma$, and chosen $\Gamma = \Gamma_L + \Gamma_R = 1$ as an 
energy unit. The asymmetry in the couplings to the leads is defined as 
$\delta = \Gamma_L/\Gamma_R$. Note that in our previous paper \cite{MK_1}, to 
faithfully reflect the situation in experiment \cite{Simmel}, we kept 
$\Gamma_R = 1$ and varied $\Gamma_L$, so $\Gamma_L + \Gamma_R \neq 1$. Here we 
shall concentrate on scaling of various physical quantities with asymmetry 
parameter $\delta$, therefore we keep $\Gamma = 1$, which means that both 
coupling parameters $\Gamma_L$ and $\Gamma_R$ change at the same time.

In the linear regime, i.e. for small voltage biases 
($eV = \mu_L - \mu_R \rightarrow 0$) and small temperature gradients 
($\delta T = T_L - T_R \rightarrow 0$) one defines the conductance 
$G = -(e^2/T) L_{11}$ , thermopower $S = -(1/eT) (L_{12}/L_{11})$ and thermal
conductance $\kappa = (1/T^2) (L_{22} - L^2_{12}/L_{11})$. The kinetic
coefficients read
\begin{eqnarray}
L_{11} = \frac{T}{h} \sum_{\sigma} \int d\omega \tilde \Gamma(\omega)
{\rm Im} G^r_{\sigma}(\omega)
\left(-\frac{\partial f(\omega)}{\partial \omega} \right)_T
\label{L11}
\end{eqnarray}
\begin{eqnarray}
L_{12} = \frac{T^2}{h} \sum_{\sigma} \int d\omega \tilde \Gamma(\omega)
{\rm Im} G^r_{\sigma}(\omega)
\left(\frac{\partial f(\omega)}{\partial T} \right)_{\mu}
\label{L12}
\end{eqnarray}
\begin{eqnarray}
L_{22} = \frac{T^2}{h} \sum_{\sigma} \int d\omega \tilde \Gamma(\omega)
(\omega - eV) {\rm Im} G^r_{\sigma}(\omega)
\left(\frac{\partial f(\omega)}{\partial T} \right)_{\mu}
\label{L22}
\end{eqnarray}
with equilibrium Fermi distribution function 
$f_L(\omega) = f_R(\omega) = f(\omega)$.

In the nonlinear regime, i.e. for any voltages and temperature differences, it
is not possible to use kinetic coefficients, as they have been derived for 
small deviations from equilibrium \cite{Callen}. Therefore, in general case, 
the thermoelectric quantities have to be calculated from proper general 
definitions, like the nonlinear differential conductance $G(eV) = dJ_e/d(eV)$, 
thermopower $S = \Delta V/\Delta T |_{J_e = 0}$ or thermal conductance 
$\kappa = - J_Q/\Delta T |_{J_e =0}$.

%%%%%%%%%%%%%%%%%%%%%%%%%%%%%%%%%%%%%%%%%%%%%%%%%%%%%%%%%%%%%%%%%%%%%%%%%%%%%%

\section{\label{linear} Linear regime}

In the linear regime the transport coefficients depend on the asymmetry
parameter via $\tilde \Gamma$ only. This stems from the fact that the
equilibrium density of states (DOS) entering Eqs. (\ref{L11})-(\ref{L22}) is
independent of asymmetry as long as $\Gamma_L + \Gamma_R = 1$. As a result 
$L_{ij}(\delta)$ can be easily obtained from its behavior in the case of 
symmetric couplings ($\delta = 1$) as 
\begin{eqnarray}
L_{ij}(\delta) = 4 L^s_{ij} \frac{\delta}{(1 + \delta)^2}, 
\label{L_scall}
\end{eqnarray}
where by $L^s_{ij}$ we denoted the corresponding kinetic coefficient for 
symmetrically coupled QD, i.e. $L^s_{ij} = L_{ij}(\delta = 1)$.

Knowing $\delta$ dependence of the $L_{ij}$, one can deduce what will be the 
modifications of the transport properties due to the asymmetry in the 
couplings. For example, the linear conductance is directly related to the
kinetic coefficient $L_{11}$ via the previously mentioned relation 
$G = -(e^2/T) L_{11}$. Thus $G(\delta)$   shows the same scaling with
$\delta$ as $L_{ij}$ does. It is linear in $\delta$ in the limit $\delta<<1$ 
and goes as $1/\delta$ for $\delta>>1$.

\begin{figure}[h]
 \resizebox{0.9\linewidth}{!}{
  \includegraphics{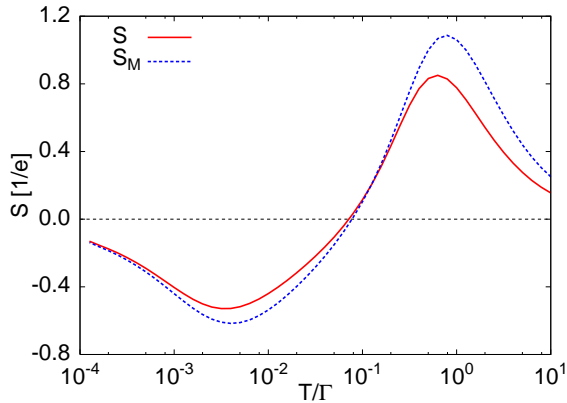}}
 \caption{\label{Fig3} The linear thermopower $S$ (solid line) and the Mott
          thermopower $S_M$ (dashed line) as a function of temperature. Both
	  quantities are not sensitive to the asymmetry $\delta$.}
\end{figure}

We shall not present the numerical results for conductance and thermal
conductance as for $\delta = 1$ as they are well known from previous studies 
\cite{Boese,Dong,MK_2}. The same is true for thermopower. However in Fig.
\ref{Fig3} we show the temperature dependence of the ($\delta$ independent)
thermopower calculated from kinetic coefficients $L_{ij}$ (solid line) and
compare it to that obtained from so called Mott formula \cite{Mott} (dashed 
line). 

Thermopower is very useful measure of the Kondo correlations 
\cite{Boese,Dong,MK_2,Costi_1}. Below the Kondo temperature $T_K$ it is 
negative, indicating particle like transport and changes sign at higher 
temperatures, where the transport is hole-like. At $T \approx \Gamma$ it shows 
a broad maximum associated with single particle excitations. The sign change 
can also be understood from the fact that $S$ is sensitive to the slope of the 
density of states at the Fermi energy. With decreasing temperature, the 
Kondo correlations lead to development of a narrow peak in the DOS slightly 
above the Fermi energy, and thus to a slope change at the $E_F$.

At this point we would like to comment on the validity of so-called Mott 
formula for thermopower \cite{Mott} , which is widely used in the literature
\cite{Scheibner,Small,Heremans}, while explaining experimental results. This 
formula relates thermopower and the linear conductance, i.e. 
\begin{eqnarray}
S_M = -\frac{\pi^2 k^2_B}{3e} \frac{T}{G} \frac{\partial G}{\partial E_F}
\label{Mott}
\end{eqnarray}
and is expected to work well in noninteracting systems at low temperatures
\cite{Lunde}. As we can see in Fig. \ref{Fig3} the Mott formula works 
surprisingly well in the interacting systems in the whole temperature range. 
The qualitative behavior of $S_M$ is approximately the same as of linear 
thermopower $S$ (solid line), giving slightly different temperature at which 
$S_M$ changes sign. It also does not depend on the asymmetry $\delta$. Thus one 
can conclude that the interactions do not necessarily lead to the violation of 
the Mott formula. Similar conclusions have been recently obtained for 
interacting quantum wires \cite{Lunde_1}. 

The low temperature thermopower has been measured in the Kondo regime in 
quantum dot system by Scheibner {\it et al.} \cite{Scheibner}. These authors 
found departures from the Mott formula and attributed them to developement of
the narrow Kondo resonance. On the other hand, the measurements of thermopower
in carbon nanotubes \cite{Small} and in Zn nanowires \cite{Heremans}
qualitatively agree with the Mott formula, Eq. (\ref{Mott}). 

\begin{figure}[h]
 \resizebox{0.9\linewidth}{!}{
  \includegraphics{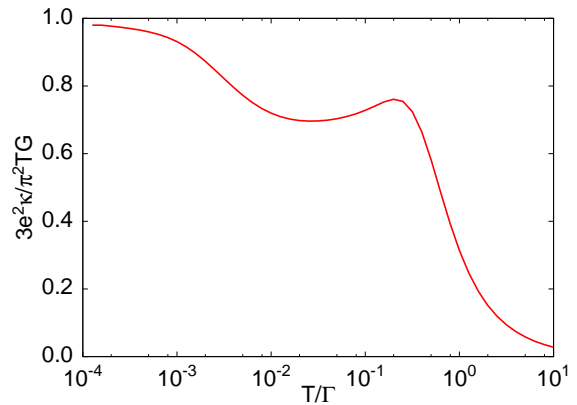}}
 \caption{\label{Fig4} The Wiedemann-Franz ratio $3 e^2/\pi^2 (\kappa/TG)$
          as a function of temperature. This quantity is not sensitive to the 
	  asymmetry $\delta$.}
\end{figure}

Figure \ref{Fig4} shows temperature dependence of the Wiedemann-Franz (WF) law 
which relates thermal and electrical transport via relation
\begin{eqnarray}
\kappa = \frac{\pi^2}{3 e^2} T G
\label{WF}
\end{eqnarray}
This law describes transport in Fermi liquid bulk metals and in general is not
obeyed in QD systems where the transport takes place through a small confined
region \cite{Boese,Dong,MK_2}. However, at very low temperatures, where the
Kondo effect develops and the ground state of the system has Fermi liquid
nature, the WF law is recovered. At high temperatures, the WF law is violated
as the transport is due to sequential processes leading to the suppression of 
the thermal transport \cite{MK_2}. The WF relation does not depend on the
asymmetry $\delta$, as both   electrical and thermal 
conductances show similar scaling with $\delta$. 

At this point we would like to comment on the validity of NCA, as it is well
known, that this approach gives non-Fermi liquid ground state 
\cite{Bickers,Cox}. However, around the Kondo temperature and slightly below
$T_K$ it provides a proper description. In our case 
$T_K = 5.3 \cdot 10^{-2} \Gamma$, thus NCA is believed to give reliable results 
in the temperature range studied here. However, at lower temperatures we expect 
a violation of WF law.

\begin{figure}[h]
 \resizebox{0.9\linewidth}{!}{
  \includegraphics{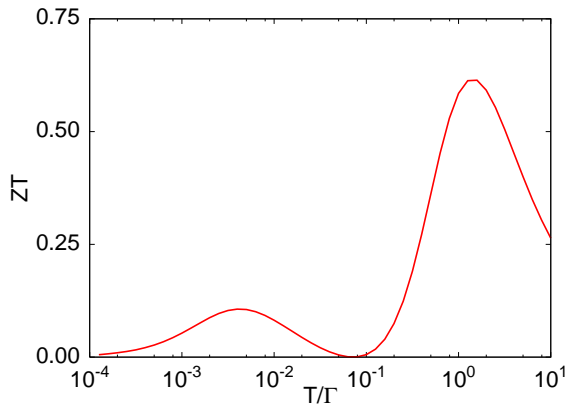}}
 \caption{\label{Fig5} Temperature dependence of the thermoelectric figure of
          merit $Z T = S^2 G T /\kappa$. This quantity does not depend on the 
	  asymmetry $\delta$.}
\end{figure}

Thermoelectric figure of merit $Z = S^2 G/\kappa$ is a direct measure of the 
usefulness of the system for applications. For simple systems it is inversely
proportional to operation temperature, thus it is more convenient to plot $ZT$,
which numerical value is an indicator of the system performance. Figure 
\ref{Fig5} shows $ZT$ as a function of temperature.
It has two maxima, one associated with charge fluctuations around $T = \Gamma$ 
and the other one associated with the Kondo correlations below the Kondo 
temperature. At temperature at which thermopower $S$ changes sign it vanishes. 
However, as one can see, its value never exceeds 1, which indicates limited 
practical applicability of the system for cryogenic purposes. Similarly as the 
Wiedemann-Franz relation, it also does not depend on the asymmetry in the 
couplings.

%%%%%%%%%%%%%%%%%%%%%%%%%%%%%%%%%%%%%%%%%%%%%%%%%%%%%%%%%%%%%%%%%%%%%%%%%%%%%%

\section{\label{nonlinear} Nonlinear transport}

In nonlinear regime we do not expect simple scaling of thermoelectric 
quantities with the asymmetry $\delta$, as the $\delta$ will modify not only 
the effective coupling $\tilde \Gamma$ in Eqs. (\ref{p_curr}) and 
(\ref{h_curr}) but also the QD density of states. In general, under 
nonequilibrium conditions the Kondo resonance in the DOS will be split by the 
bias voltage $eV = \mu_L - \mu_R$. To be precise, there will be two resonances, 
one located at $\omega = \mu_L$ and the other one located at $\omega = \mu_R$. 
If we introduce asymmetry in the couplings $\delta$, one of the resonances will 
be broader and higher and the other one will be narrower and lower, depending 
which of the couplings $\Gamma_L$ or $\Gamma_R$ is stronger \cite{MK_1}. As a 
results, all the thermoelectric quantities will be modified in more complicated 
way.

\begin{figure}[h]
 \resizebox{0.9\linewidth}{!}{
  \includegraphics{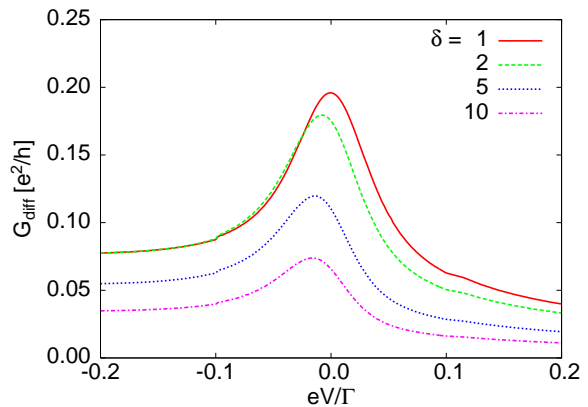}}
 \caption{\label{Fig7} Differential conductance $G_{\text{diff}}$ as a function 
          of the source-drain voltage $eV$ for different coupling asymmetries
	  $\delta = 1$, $2$, $5$ and $10$ (from top to bottom), calculated at 
	  temperature $T = 10^{-2} \Gamma$. Note evolution of the position of 
	  the resonance with $\delta$.}
\end{figure}

\begin{figure}[h]
 \resizebox{0.9\linewidth}{!}{
  \includegraphics{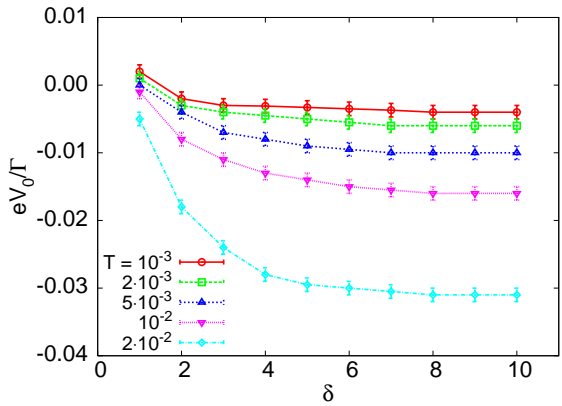}}
 \caption{\label{Fig8} The position of maximum in differential conductance as a
          function of the asymmetry parameter $\delta$ for various temperatures 
	  $T$. Each position of the maximum has been extracted from
	  corresponding $dJ_e/d(eV)$ vs. $eV$ characteristics with uncertainty
	  given by errorbars.}
\end{figure}

The most pronounced example is the differential conductance 
$G_{\text{diff}} = dJ_e/d(eV)$. In symmetrically coupled QD there is a 
resonance at zero source-drain bias. The introduction of asymmetry in the 
couplings ($\delta \neq 1$) shifts the resonance to non-zero voltage 
\cite{MK_1}. Figure \ref{Fig7} shows differential conductance $G_{\text{diff}}$ 
for different values of the asymmetry parameter $\delta$. 
The spectra are calculated for temperature $T = 10^{-2} \Gamma$, below the 
Kondo temperature $T_K$, which is equal to $5.3 \cdot 10^{-2} \Gamma$ in this 
case. One observes the evolution of the position of the maximum towards 
negative voltages with increasing of the $\delta$. This is the effect of 
different heights and widths of two Kondo resonances in the density of states, 
one at $\mu_L$ and the other one at $\mu_R$ \cite{MK_1}. It is also clearly 
seen that the asymmetry suppresses the conductance for all bias voltages. This 
is due to suppression of the effective coupling $\tilde \Gamma$ in Eq. 
(\ref{p_curr}) by the asymmetry parameter $\delta$. At lower temperatures the
maximum of the conductance is narrower due to temperature effects and similar 
suppression of the conductance with increasing $\delta$ is observed. In Fig. 
\ref{Fig8}, the position of the conductance maximum $eV_0$ is plotted as a 
function of the asymmetry parameter $\delta$ for various temperatures. 
The reason for larger shifts of the maxima at higher temperatures stems from
two facts. First one is the asymmetry in the couplings, which leads to a 
broader and higher (narrower and lower) Kondo resonance associated with
tunneling between QD and left (right) lead. The second reason, more important,
is the temperature effect. At very low temperature, the energy window 
accessible to the transport is bounded by the positions of the chemical 
potentials $\mu_L$ and $\mu_R$. So in fact, only half of each Kondo resonance 
contributes to transport. At higher temperatures, this energy window becomes 
wider, thus it increases effect of the asymmetry, leading to larger shifts of 
the maxima in $G_{\text{diff}}$. All curves in Fig. \ref{Fig8} show  
$f(\delta) = a + b/\delta$ behavior  with temperature dependent coefficients 
$a$ and $b$. 

\begin{figure}[h]
 \resizebox{0.9\linewidth}{!}{
  \includegraphics{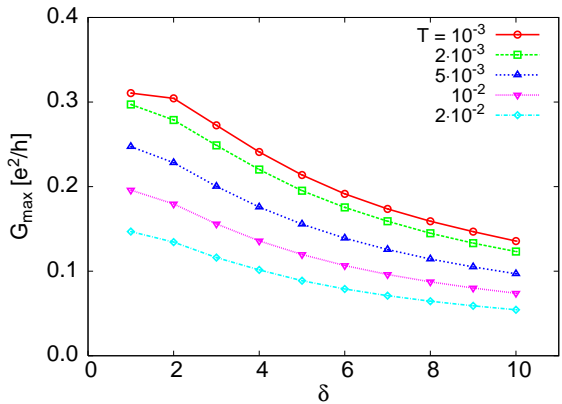}}
 \caption{\label{Fig9} Peak height in $G_{\text{diff}}$ vs. asymmetry parameter
          $\delta$ for a number of temperatures $T$.}
\end{figure}

Similarly, the maximal value  of the differential conductance (associated with 
the Kondo effect), i.e $G_{\text{diff}}(eV_0)$, also depends on the asymmetry 
parameter $\delta$, as  shown in Fig. \ref{Fig9}. 

\begin{figure}[h]
 \resizebox{0.9\linewidth}{!}{
  \includegraphics{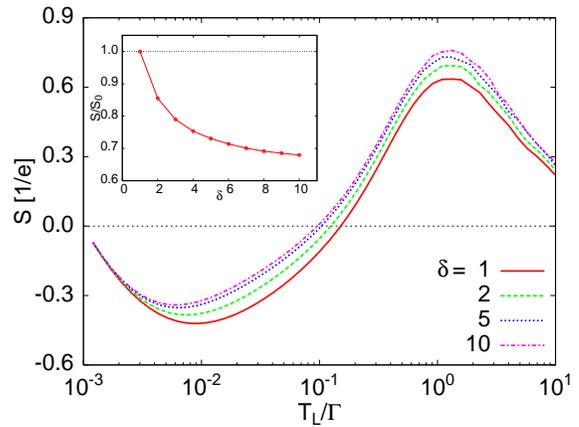}}
 \caption{\label{Fig10} Nonlinear thermopower $S$ as a function of the left 
          lead temperature $T_L$ for various values of the asymmetry parameter 
	  $\delta$. The temperature of the right electrode $T_R$ is fixed and
	  equal to $10^{-3} \Gamma$. Note that, unlike linear thermopower (Fig. 
	  \ref{Fig3}), the nonlinear $S$ strongly depends on $\delta$. Inset:
	  Normalized thermopower ($S(\delta)/S(\delta = 1)$ as a function of 
	  the asymmetry $\delta$ calculated for 
	  $T_L = 1.9 \cdot 10^{-2} \Gamma$.}
\end{figure}

Another quantity affected by the asymmetry in the couplings is the nonlinear
thermopower, shown in Fig. \ref{Fig10} as a function of the left lead
temperature $T_L$. 
The right lead is kept at fixed temperature equal to $10^{-3} \Gamma$. 
Similarly as in the linear regime, the thermopower is negative in the low 
temperature (Kondo) regime and positive at high temperatures. The asymmetry 
$\delta$ leads to shallower minimum of $S$ in the Kondo regime, and to higher 
values of $S$ at temperature around $\Gamma$, associated with single particle 
excitations. The change of the minimum of $S$ at low temperatures reflects a 
fact that asymmetry changes the slope of the QD density of states at the Fermi 
energy due to splitting of the Kondo resonance under nonequilibrium. There is 
no simple scaling behavior of nonlinear thermopower with asymmetry $\delta$. 

Another interesting fact is that $S$ changes sign at different temperatures 
for different asymmetries $\delta$, leading to different Kondo temperatures. At 
first sight this is in contradiction with the conclusions reached from the 
linear behavior of the thermopower. However one has to keep in mind that for 
the system out of equilibrium  ($\mu_L \neq \mu_R$ and $T_L \neq T_R$) one of 
the Kondo peaks may disappear. For the parameters studied we always have the 
Kondo effect associated with the right lead but not with the left one. The 
density of states has a Kondo resonance at the energy $\omega = \mu_R$, while 
the Kondo resonance at $\mu_L$ vanishes with increasing $T_L$. If the asymmetry 
in the couplings is introduced, the  Kondo resonance at $\mu_R$ contributes 
less and less to electrical and thermal transport. As a result, for large 
asymmetry $\delta$ one observes lower minimum at low $T_L$ and larger values of 
$S$ at higher temperatures. 

The nonlinear thermal conductance is not sensitive to the Kondo correlations 
due to small contribution of low energy excitations to it. As a result the 
nonlinear thermal conductance $\kappa$ shows similar scaling behavior with 
$\delta$ as the linear one. In other words, the nonlinear $\kappa$ can be 
calculated, with good accuracy, from its linear values.

It is the nonlinear differential conductance which changes with $\delta$ in 
qualitative way. In symmetrically coupled QD ($\delta = 1$) it shows a maximum 
at zero bias, while the asymmetry tends to move the maximum to non-zero 
voltages. All the other quantities change with $\delta$ only quantitatively.   

%%%%%%%%%%%%%%%%%%%%%%%%%%%%%%%%%%%%%%%%%%%%%%%%%%%%%%%%%%%%%%%%%%%%%%%%%%%%%%

\section{\label{conclusions} Conclusions}

We have studied electrical and thermal transport in a quantum dot 
asymmetrically coupled to external leads. Those studies revealed that the most 
suitable tool to study the asymmetry in the couplings $\delta$ is nonlinear 
differential conductance. The presence of the maximum at non-zero bias voltage
directly signals that the system is asymmetrically coupled to the electrodes. 
In the linear regime electrical and thermal conductances show simple scaling 
behavior with $\delta$ while thermopower does not depend on it. Finally, we 
also checked the Mott formula of thermopower and found that this quantity is a 
good approximation even in the presence of strong Coulomb interactions, 
provided the changes in the spectrum are confined to small region around 
chemical potential.

\begin{acknowledgements}
This work has been supported by the KBN grant no. 1 P03B 004 28 (MK) and 
PBZ-MIN-008/P03/2003 (KIW).
\end{acknowledgements}

%%%%%%%%%%%%%%%%%%%%%%%%%%%%%%%%%%%%%%%%%%%%%%%%%%%%%%%%%%%%%%%%%%%%%%%%%%%%%%

\end{document}